\documentclass{appolb}
\usepackage{graphicx}
\usepackage{verbatim}
\usepackage{color}
\usepackage{cite}

\begin{document}

\title{Upper limits for the production of the $\eta$-mesic Helium in the $dd\rightarrow$ $^{3}\hspace{-0.03cm}\mbox{He} n \pi{}^{0}$ and $dd\rightarrow$ $^{3}\hspace{-0.03cm}\mbox{He} p \pi{}^{-}$ reactions%
\thanks{Jagiellonian Symposium on Fundamental and Applied Subatomic Physics}%
}
\author{Magdalena Skurzok$^{a}$, Wojciech Krzemie{\'n}$^{b}$, Oleksandr Rundel$^{a}$ and Pawel Moskal$^{a}$\\
\small{FOR THE WASA AT COSY COLLABORATION}
\address{$^{a}$M. Smoluchowski Institute of Physics, Jagiellonian University, Cracow, Poland\\
}
\address{$^{b}$National Centre for Nuclear Research, {\'S}wierk, Poland}
}

\maketitle
\begin{abstract}

We performed a search for $^{4}\hspace{-0.03cm}\mbox{He}$-$\eta$ bound state in $dd\rightarrow$ $^{3}\hspace{-0.03cm}\mbox{He} n \pi{}^{0}$ and $dd\rightarrow$ $^{3}\hspace{-0.03cm}\mbox{He} p \pi{}^{-}$ reactions with the WASA-at-COSY facility using a ramped beam technique. The measurement was carried out with high statistics and high acceptance.~The signature of $\eta$-mesic nuclei was searched for by the measurement of the excitation functions in the vicinity of the $\eta$ production threshold for each of the considered channels.~We did not observe the narrow structure which could be interpreted as a bound state. The preliminary upper limits of the total cross sections for the bound state production and decay varies from 21~nb to 36~nb for the $dd\rightarrow$ $^{3}\hspace{-0.03cm}\mbox{He} n \pi{}^{0}$ channel, and from 5~nb to 9~nb for the $dd\rightarrow$ $^{3}\hspace{-0.03cm}\mbox{He} p \pi{}^{-}$ channel for the bound state width ranging from 5 to 50~MeV.


\end{abstract}

  

\section{Introduction}

Since Haider and Liu postulated a possible existence of $\eta$-mesic nuclei~\cite{HaiderLiu1}, many experimental groups performed measurements dedicated to search for the new kind of nuclear matter in which the $\eta$ meson is bound within a nucleus via the strong interaction. However, till now, none of the experiments have brought the clear evidence for the bound state existence. The status of the search was recently described in the following reviews~\cite{Machner_2015,Kelkar,Kelkar_new,Haider_new,Krusche_Wilkin,Bass_Moskal,Moskal_2016}. Some of the experiments set the upper limits for the bound state production in several processes. COSY-11~\mbox{~\cite{Smyrski1,Krzemien1,MoskalSmyrski}} group estimated the upper limit of total cross section for $dp\rightarrow$ \mbox{$(^{3}\hspace{-0.03cm}\mbox{He}$-$\eta)_{bound} \rightarrow ppp\pi^{-}$} process to the value of 270~nb and for $dp\rightarrow$ $(^{3}\hspace{-0.03cm}\mbox{He}$-$\eta)_{bound} \rightarrow$ $^{3}\hspace{-0.03cm}\mbox{He}\pi^{0}$ to the value 70~nb. COSY-GEM measurement of $p^{27}\hspace{-0.03cm}Al\rightarrow$ $^{3}\hspace{-0.03cm}\mbox{He}p\pi^{-}X$ brought the upper limit of the total cross section for $(^{25}\hspace{-0.03cm}\mbox{Mg}$-$\eta)_{bound}$ production equal to 0.46 $\pm$ 0.16(stat) $\pm$ 0.06(syst)~nb~\cite{Budzanowski}. The WASA measurement in 2008 results in the upper limit of of the total cross section for the $(^{4}\hspace{-0.03cm}\mbox{He}$-$\eta)_{bound}$ creation in $dd\rightarrow$ $^{3}\hspace{-0.03cm}\mbox{He}p\pi^{-}$ reaction, which varies from 20~nb to 27~nb for the range of the bound state width from 5~MeV to 35~MeV~\cite{Adlarson_2013,Krzemien_PhD}. The measurement carried out two years later permitted to lower the upper bound for the cross section of $dd\rightarrow(^{4}\hspace{-0.03cm}\mbox{He}$-$\eta)_{bound}\rightarrow$ $^{3}\hspace{-0.03cm}\mbox{He} p \pi{}^{-}$ process down to the value of few nanobarns. Additionally, the upper limit of the preliminary total cross section was determined for the first time for the $(^{4}\hspace{-0.03cm}\mbox{He}$-$\eta)_{bound}$ production in $dd\rightarrow$ $^{3}\hspace{-0.03cm}\mbox{He}n\pi^{0}$ reaction~\cite{MSkurzok_PhD}. This paper presents the preliminary results obtained for the aforementioned  processes.

\section{Experimental results}

In November 2010, WASA-at-COSY Collaboration carried out the experiment dedicated for the search for $^{4}\hspace{-0.03cm}\mbox{He}$-$\eta$ bound states in $dd\rightarrow$ $^{3}\hspace{-0.03cm}\mbox{He} n \pi{}^{0}$  and $dd\rightarrow$ $^{3}\hspace{-0.03cm}\mbox{He} p \pi{}^{-}$ reactions. The ramped beam technique was used to vary the momentum continuously from 2.127~GeV/c to 2.422~GeV/c, which corresponds to a range of excess energies $Q$ from -70 to 30~MeV~\cite{MSkurzok, WKrzemien_2015}. The detailed description of the WASA experimental setup is presented in~\cite{WASA_dsc}.


Analysis for the $dd\rightarrow$ $^{3}\hspace{-0.03cm}\mbox{He} n \pi{}^{0}$ and $dd\rightarrow$ $^{3}\hspace{-0.03cm}\mbox{He} p \pi{}^{-}$ reactions were carried out independently. Next, the set of the cross-check tests was performed to assure the consistency at the PID level. The $^{3}\hspace{-0.03cm}\mbox{He}$ ions and nucleon-pion pairs were identified in the Forward and Central Detector, respectively. The deposited energy patterns in thick scintillator layers of the Forward Hodoscope was used to identify the $^{3}\hspace{-0.03cm}\mbox{He}$ ions (the \mbox{$\Delta$E-E method}).~The neutral pion $\pi^{0}$ was reconstructed based on the invariant mass of two gamma quanta while the neutron was identified via the missing mass technique~\cite{MSkurzok_PhD}. The proton and $\pi^{-}$ identification was based on the measurement of the energy loss in the thin Plastic Scintillator Barrel combined with the energy deposited in the Electromagnetic Calorimeter~\cite{Adlarson_2013}. 

\begin{figure}[h!]
\centering
\includegraphics[width=6.0cm,height=4.0cm]{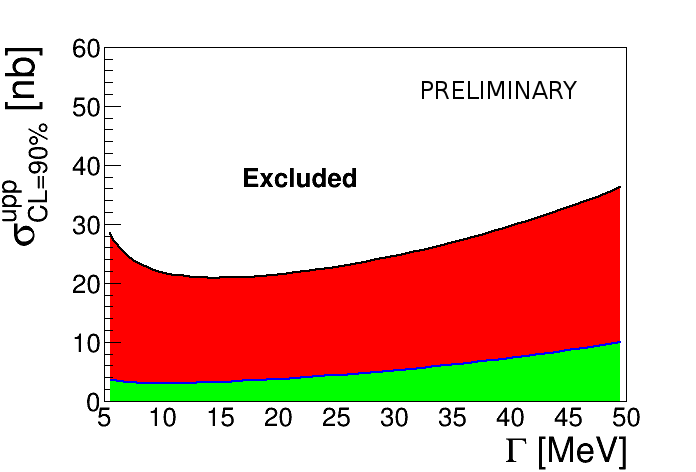}
\includegraphics[width=6.0cm,height=4.0cm]{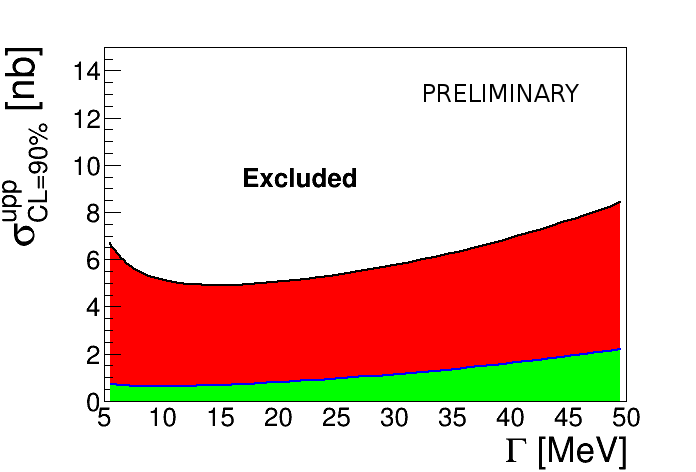}
\caption{Preliminary upper limit of the total cross-section for $dd\rightarrow(^{4}\hspace{-0.03cm}\mbox{He}$-$\eta)_{bound}\rightarrow$ $^{3}\hspace{-0.03cm}\mbox{He} n \pi{}^{0}$ (left panel) and $dd\rightarrow(^{4}\hspace{-0.03cm}\mbox{He}$-$\eta)_{bound}\rightarrow$ $^{3}\hspace{-0.03cm}\mbox{He} p \pi{}^{-}$ (right panel) reaction as a function of the width of the bound state. The binding energy was set to 30~MeV. The green areas denote the systematic uncertainties~\cite{MSkurzok_PhD}.~\label{Result_sigma_upp}}  
\end{figure}

The events which may correspond to the bound states production were selected using criteria based on Monte Carlo simulations for the $\eta$-mesic nuclei production and decay. We apply the cuts in the momentum of $^{3}\hspace{-0.03cm}\mbox{He}$ in the CM frame, nucleon CM kinetic energy, pion CM kinetic energy and the opening angle between nucleon-pion pair in the CM. The region rich in signal corresponds to the momenta of the $^{3}\hspace{-0.03cm}\mbox{He}$ in the range \mbox{$p^{cm}_{^{3}\hspace{-0.05cm}He}\in(0.07,0.2)$~GeV/c}. For this region the excitation function was obtained by normalizing the events selected in individual excess energy intervals by the corresponding integrated luminosities (the detailed description of the luminosity determination one can find in Ref.~\cite{MSkurzok_PhD,MSkurzok_2015}) and corrected for acceptance and efficiency. The excitation function does not reveal the resonance-like structure, which could be the signature of the $\eta$-mesic nuclei existence~\cite{MSkurzok_PhD}, however the interpretation of the results is still in progress. So far, the upper limit of the total cross section for the $dd\rightarrow(^{4}\hspace{-0.03cm}\mbox{He}$-$\eta)_{bound}\rightarrow$ $^{3}\hspace{-0.03cm}\mbox{He} n \pi{}^{0}$ and $dd\rightarrow(^{4}\hspace{-0.03cm}\mbox{He}$-$\eta)_{bound}\rightarrow$ $^{3}\hspace{-0.03cm}\mbox{He} p \pi{}^{-}$ processes was determined on the 90\% confidence level. Preliminary, the upper limits were obtained by the fit of the sum of the polynomial and Breit-Wigner functions to the experimentally determined excitation functions. It varies from 21 to 36 nb for the first channel and from 5 to 9 nb for the second channel for the bound state width ranging from 5 to 50 MeV (See Fig.~\ref{Result_sigma_upp}). 

\vspace{-0.3cm}

\begin{figure}[h!]
\centering
\includegraphics[width=6.0cm,height=4.0cm]{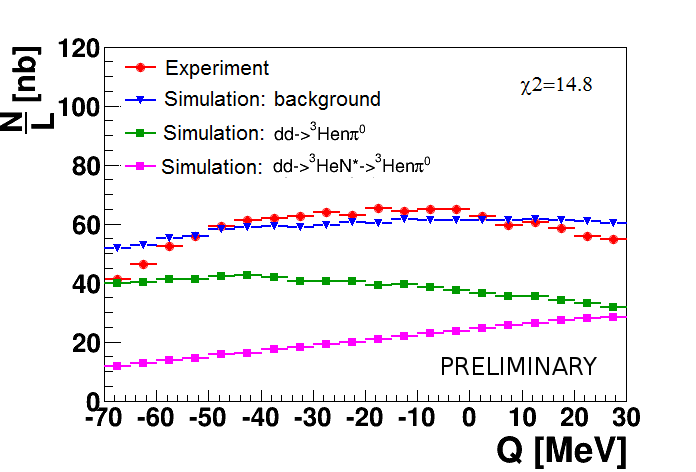}
\includegraphics[width=6.0cm,height=4.0cm]{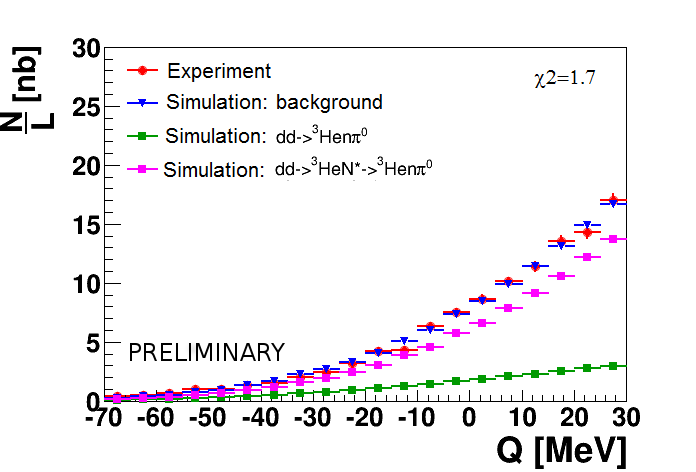}
\caption{Preliminary experimental excitation functions (red circles) fitted with two background reactions: $dd\rightarrow$ $^{3}\hspace{-0.03cm}\mbox{He} n \pi{}^{0}$ (green squares) and $dd\rightarrow ^{3}\hspace{-0.03cm}\mbox{He}N^{*}$ $\rightarrow$ $^{3}\hspace{-0.03cm}\mbox{He} n \pi{}^{0}$ (magenta squares). A sum of both background contributions is shown as blue triangles. Left and right panels show results for the regions rich in signal and poor in signal, respectively. The figure is adopted from~\cite{MSkurzok_PhD}.~\label{fit_plots}}
\end{figure}

A possible broad state in the case of $dd\rightarrow(^{4}\hspace{-0.03cm}\mbox{He}$-$\eta)_{bound}\rightarrow$ $^{3}\hspace{-0.03cm}\mbox{He} n \pi{}^{0}$ reaction cannot be excluded by the current data set~\cite{MSkurzok_PhD}. The kinematic region, where we expect the evidence of the signal from the bound state corresponding to $^{3}\hspace{-0.03cm}\mbox{He}$ momenta in the CM system in range \mbox{$p^{cm}_{^{3}\hspace{-0.05cm}He}\in(0.3,0.4)$~GeV/c}, cannot be fully described only by the combination of the considered background processes (see left panel of Fig.~\ref{fit_plots}). In contrast, as shown in the right panel of Fig.~\ref{fit_plots} the experimental excitation function is very well fitted by the background contributions for the region where the signal is not expected. 

\section{Conclusion and Perspectives}

The excitation functions were determined for $dd\rightarrow(^{4}\hspace{-0.03cm}\mbox{He}$-$\eta)_{bound}\rightarrow$ $^{3}\hspace{-0.03cm}\mbox{He} p \pi{}^{-}$ and $dd\rightarrow(^{4}\hspace{-0.03cm}\mbox{He}$-$\eta)_{bound}\rightarrow$ $^{3}\hspace{-0.03cm}\mbox{He} n \pi{}^{0}$ processes, however none of them reveal any direct narrow structure which could be signature of the bound state with width less than 50~MeV. The interpretation of the resuts is still in progress. So far preliminary upper limit of the total cross section for the $\eta$-mesic $^{4}\hspace{-0.03cm}\mbox{He}$ formation and decay was estimated. In case of $dd\rightarrow(^{4}\hspace{-0.03cm}\mbox{He}$-$\eta)_{bound}\rightarrow$ $^{3}\hspace{-0.03cm}\mbox{He} p \pi{}^{-}$ reaction we obtained the preliminary upper limit of the total cross section in order of few nb which is about four times lower in comparison with the result obtained from 2008 data~\cite{Adlarson_2013}. Comparing to theoretically estimated value~\cite{WycechKrzemien}, the obtained upper limit value does not exclude the existence of the bound state. The excitation function for the reaction $dd\rightarrow(^{4}\hspace{-0.03cm}\mbox{He}$-$\eta)_{bound}\rightarrow$ $^{3}\hspace{-0.03cm}\mbox{He} n \pi{}^{0}$ was obtained for the first time in the experiment. The obtained upper limit is here by factor of five larger than predicted value therefore, we can conclude, that the current measurement does not exclude the existence of bound state also in this process~\cite{Kelkar_2015_new}. Moreover, the excitation function obtained for this reaction is a subject of interpretation of few theoretical groups\footnote{N. Kelkar and S Hirenzaki (Presentations at the Jagiellonian Symposium on Fundamental and Applied Subatomic Physics, Cracow).} with respect to very wide \mbox{$(^{4}\hspace{-0.03cm}\mbox{He}$-$\eta)_{bound}$} or \mbox{$^{3}\hspace{-0.03cm}\mbox{He}$-$N^{*}$} bound state~\cite{Kelkar_2015_new}.


\indent In May 2014, we extended the search for to $^{3}\hspace{-0.03cm}\mbox{He}$-$\eta$ sector~\cite{Proposal_2014}. We chose processes corresponding to the three mechanisms: (i) absorption of the $\eta$ meson by one of the nucleons, which subsequently decays into $N^{*}$-$\pi$ pair e.g.: $pd \rightarrow$ ($^{3}\hspace{-0.03cm}\mbox{He}$-$\eta)_{bound} \rightarrow$ $p p p \pi{}^{-}$ , (ii) decay of the $\eta$ -meson while it is still "orbiting" around a nucleus e.g.: $pd \rightarrow$ ($^{3}\hspace{-0.03cm}\mbox{He}$-$\eta)_{bound} \rightarrow$ $^{3}\hspace{-0.03cm}\mbox{He} 6\gamma$ or $pd \rightarrow$ ($^{3}\hspace{-0.03cm}\mbox{He}$-$\eta)_{bound} \rightarrow$ $^{3}\hspace{-0.03cm}\mbox{He} 2\gamma$ reactions and (iii) $\eta$ meson absorption by few nucleons e.g.: $pd \rightarrow$ ($^{3}\hspace{-0.03cm}\mbox{He}$-$\eta)_{bound} \rightarrow$ $ppn$ or $pd \rightarrow$ ($^{3}\hspace{-0.03cm}\mbox{He}$-$\eta)_{bound} \rightarrow$ $pd$. Almost two weeks of measurement with an average luminosity of about 6$\cdot10^{30}$ cm$^{-2}$ s$^{-1}$ allowed to collect a world largest data sample for $^{3}\hspace{-0.03cm}\mbox{He}$-$\eta$. The data analysis is in progress.

The search for $\eta$ and $\eta'$ - mesic bound states is carried out also by other international collaborations, e.g. at J-PARC~\cite{Fujioka1, Fujioka2} and at GSI~\cite{Yoshiki, Fujioka3}. In parallel, several theoretical studies are ongoing~\cite{ WycechKrzemien, BassTom1, Hirenzaki_2010, Hirenzaki1, Friedman_2013, Wilkin2, Nagahiro_2013, Kelkar, Niskanen_2015, Wilkin_2016}. 


\section{Acknowledgements}

\noindent We acknowledge support by the Foundation for Polish Science - MPD program, co-financed by the European
Union within the European Regional Development Fund, by the Polish National Science Center through grants No.~DEC-2013/11/N/ST2/04152, 2011/01/B/ST2/00431, 2011/03/B/ST2/ 01847 and by the FFE grants of the Forschungszentrum J\"ulich.

\addcontentsline{toc}{chapter}{Bibliography}

\thispagestyle{plain}



\begin{thebibliography}{99}


\bibitem{HaiderLiu1} Q.~Haider, L. C.~Liu, \textit{Phys. Lett.} \textbf{B172}, 257 (1986). 

\bibitem{Machner_2015} H. Machner, \textit{J. Phys.}, \textbf{G42}, 043001 (2015).

\bibitem{Kelkar} N. G. Kelkar \textit{et al., Rept. Prog. Phys.} \textbf{76}, 066301 (2013).

\bibitem{Kelkar_new} N. G. Kelkar, \textit{Acta Phys. Polon.} \textbf{B46}, 113 (2015) 1. 

\bibitem{Haider_new} Q. Haider, L.-C. Liu, \textit{Int. J. Mod. Phys.} \textbf{E24}, 1530009 (2015) 10. 

\bibitem{Krusche_Wilkin} B. Krusche, C. Wilkin, \textit{Prog. Part. Nucl. Phys.}, \textbf{80}, 43 (2014).

\bibitem{Bass_Moskal} S. Bass, P. Moskal, \textit{Acta Phys. Pol.} \textbf{B} (2016), these proceedings.

\bibitem{Moskal_2016} P. Moskal, \textit{Acta Phys. Pol.} \textbf{B} (2016) in print;  55 Cracow School of Theoretical Physics, Zakopane 2015.

\bibitem{Smyrski1} J.~Smyrski \textit{et al., Phys. Lett.} \textbf{B649}, 258 (2007). 

\bibitem{Krzemien1} W. Krzemie\'n \textit{et al., Int. J. Mod. Phys.} \textbf{A24}, 576 (2009).

\bibitem{MoskalSmyrski} P. Moskal, J. Smyrski, \textit{Acta Phys. Pol.} \textbf{B41}, 2281 (2010). 

\bibitem{Budzanowski} A. Budzanowski \textit{et al., Phys. Rev.} \textbf{C79}, 012201 (2009).

\bibitem{Adlarson_2013} P. Adlarson \textit{et al., Phys. Rev.} \textbf{C87}, 035204 (2013).

\bibitem{Krzemien_PhD} W. Krzemien, \textit{PhD Thesis, Jagiellonian Univ., arXiv:1101.3103} (2011).

\bibitem{MSkurzok_PhD} M. Skurzok, \textit{PhD Thesis, Jagiellonian Univ., arXiv:1509.01385} (2015).

\bibitem{MSkurzok} M. Skurzok, P. Moskal, W. Krzemien, \textit{Prog. Part. Nucl. Phys.} \textbf{67}, 445 (2012).

\bibitem{WKrzemien_2015} W. Krzemien, P. Moskal, M. Skurzok, \textit{Acta Phys. Pol.} \textbf{B46}, 757 (2015).

\bibitem{WASA_dsc} WASA-at-COSY Collaboration: H.-H. Adam et al., \textit{arXiv:nucl-ex/0411038} (2004).



\bibitem{MSkurzok_2015} M. Skurzok, W. Krzemien, P. Moskal, \textit{Acta Phys. Pol.} \textbf{B46}, 133 (2015).
\bibitem{WycechKrzemien} S. Wycech, W. Krzemien, \textit{Acta Phys. Pol.} \textbf{B45}, 745 (2014).

\bibitem{Kelkar_2015_new} N. G. Kelkar, D. Bedoya Ferro, P. Moskal, \textit{arXiv: 1512.01535} (2015)

\bibitem{Proposal_2014} P. Moskal, W. Krzemien, M. Skurzok, \textit{COSY proposal No. 186.3} (2014).

\bibitem{Fujioka1} H. Fujioka, \textit{LAMPF Progress Report}, \textbf{B41}, 2261 (2010).

\bibitem{Fujioka2} H. Fujioka, \textit{J. Phys. Conf. Ser.}, \textbf{374}, 012015 (2012).

\bibitem{Yoshiki} K. Yoshiki \textit{et. al. arXiv:1503.03566, Proceedings of the 20th International Conference on Particles and Nuclei (PANIC 14), DOI:10.3204/DESY-PROC-2014-04/106}, 24-29 August 2014, Hamburg, Germany, page 286, 2015.

\bibitem{Fujioka3} H. Fujioka \textit{et. al., Hyperfine Interact.} \textbf{234}, 33 (2015) 1-3.


\bibitem{BassTom1} S. D. Bass, A. W. Thomas, \textit{Acta Phys. Pol.} \textbf{B45}, 627 (2014).

\bibitem{Hirenzaki_2010} S. Hirenzaki, \textit{et al., Acta Phys. Pol.} \textbf{B41}, 2211 (2010).

\bibitem{Hirenzaki1} S. Hirenzaki, H. Nagahiro, \textit{Acta Phys. Pol.} \textbf{B45}, 619 (2014).

\bibitem{Friedman_2013} E. Friedman, A. Gal, J. Mares, \textit{Phys. Lett.} \textbf{B725}, 334 (2013).

\bibitem{Wilkin2} C.~Wilkin, \textit{Phys. Lett.} \textbf{B 654}, 92 (2007).

\bibitem{Nagahiro_2013} H. Nagahiro \textit{et al., Phys. Rev.} \textbf{C87}, 045201 (2013).

\bibitem{Niskanen_2015} J. Niskanen, \textit{Phys. Rev.} \textbf{C92}, 055205 (2015) 5. 

\bibitem{Wilkin_2016} C. Wilkin, \textit{Acta Phys. Pol.} \textbf{B} (2016), these proceedings.

\end{thebibliography}
\end{document}